\documentclass[conference]{IEEEtran}
\normalsize
\usepackage[cmex10]{amsmath}
\usepackage{amsfonts}
\usepackage{amsbsy}
\usepackage{amssymb}
\usepackage{array}
\usepackage{mdwtab}
\usepackage{graphicx}
\usepackage{cite}
\usepackage{subfigure}
\usepackage{algorithmic}
\usepackage{cases}
\usepackage{subeqnarray}
\usepackage{multirow}
\usepackage{enumerate}
\makeatletter
\def\hlinew#1{
  \noalign{\ifnum0=`}\fi\hrule \@height #1 \futurelet
   \reserved@a\@xhline}
\makeatother
\def\Var{{\rm Var}\,}
\def\E{{\rm E}\,}

\newenvironment{proof}[1][Proof]{\begin{trivlist}
\item[\hskip \labelsep {\bfseries #1}]}{\end{trivlist}}

\newcommand{\qed}{\nobreak \ifvmode \relax \else
      \ifdim\lastskip<1.5em \hskip-\lastskip
      \hskip1.5em plus0em minus0.5em \fi \nobreak
      \vrule height0.75em width0.5em depth0.25em\fi}
\title{Fine Residual Carrier Frequency and Sampling Frequency Estimation in Wireless OFDM Systems}
\author{\IEEEauthorblockN{Chen~Chen,~Yun~Chen*,~and~Xiaoyang~Zeng}
\IEEEauthorblockA{ASIC and System State Key Laboratory,~Fudan~University,~Shanghai,~China,~201203\\
Email:~chenyun@fudan.edu.cn}
}
\begin{document}
\maketitle
\begin{abstract}
This paper presents a novel algorithm for residual phase estimation in wireless OFDM systems, including the carrier frequency offset (CFO) and the sampling frequency offset (SFO). The subcarriers are partitioned into several regions which exhibit pairwise correlations. The phase increment between successive OFDM blocks is exploited which can be estimated by two estimators with different computational loads. Numerical results of estimation variance are presented. Simulations indicate performance improvement of the proposed technique over several conventional schemes in a multipath channel. 
\end{abstract}

\section{Introduction}
Although the Orthogonal-Frequency-Division-Multiplexing (OFDM) technique significantly enhances the system performance under frequency-selective fading channels, it is vulnerable to synchronization non-idealities, including the symbol timing offset (STO), carrier frequency offset (CFO), and sampling frequency offset (SFO). 

The previous works including \cite{VDB,Moose,Cox} deal with the coarse STO and CFO estimation in time domain before Fast Fourier Transform (FFT). However, due to the imperfections of compensation, after FFT, the residual part of CFO remains to be corrected. Also, at this stage, SFO should be estimated and removed; otherwise, it would lead to a phase rotation not only proportional to the tone index within one OFDM block (inter-block increment), but also grows linearly for successive OFDM blocks (intra-block increment) \cite{PolletJour}.      

In literature, several schemes are proposed to estimate or track the residual CFO and SFO in frequency domain with the assistance from pilot subcarriers \cite{SpethII,ShouYinLiu,JLS}. In \cite{SpethII}, Speth et al. utilize the symmetric locations of pilots to estimate CFO and SFO jointly. However, its performance degrades in the multipath channels. \cite{ShouYinLiu} suggests three estimators with the help of the least square estimation (LSE). An improved weighted LSE variant is proposed by Tsai et al. in \cite{JLS} which requires the second-order statistics of the channel state information (CSI). In general, these schemes mainly rely on the linearly growing inter-block increment.

This paper proposes a novel technique to make use of the intra-block increment spanning a number of OFDM blocks. By dividing the subcarrier index into several regions, the method is capable of exploiting the pairwise correlation which leads to accurate results after applying least square fitting. Two variants differing in computation complexity are presented with their numerical variances derived.            

The rest of the paper is structured as follows. Section \ref{sec:signalmodel} presents the signal model in presence of CFO and SFO. Section \ref{sec:proposed} introduces the proposed technique and analytical results of variance. Simulation results are given in Section \ref{sec:sim}. Finally, Section \ref{sec:conclusion} concludes the paper.    

\section{OFDM Signal Model with CFO and SFO} \label{sec:signalmodel}
We consider an OFDM system where the transmitted data is modulated by an $N$-point Inverse FFT (IFFT). Assuming a total of $M$ OFDM blocks to be delivered and each block consists of $K$ data samples ($K\leq N$), the complex baseband signal is described by 
\begin{equation}
s(t)=\frac{1}{\sqrt{N}}\sum_{l=0}^{M-1}\sum_{k\in \mathbb{X}}X_{l,k}e^{\frac{j2\pi k(t-(N_{g}+lN_{B})T_{s})}{NT_{s}}}\sqcap(t-lN_{B}T_{s})
\end{equation}         
where $\mathbb{X}$ are the locations of the $K$ data subcarriers; for $k \notin \mathbb{X}$, $X_{l,k}$ is either pilot or null subcarrier. $N_{g}$ is the length of the guard interval, $N_{B}$ the total length of an entire OFDM block given by $N_{B}\triangleq N+N_{g}$, and $T_{s}$ the sampling interval. $\sqcap(\cdot)$ is the rectangular function defined as
\begin{equation}\label{equ:recfunc}
\sqcap(t)=\begin{cases}
1 & 0 \leq t \leq N_{B}T_{s},\\
0 & \mathrm{otherwise}
\end{cases}
\end{equation}
 
The multipath channel is 
\begin{equation}\label{equ:channel}
h(t,\tau)=\sum_{\ell=0}^{L-1}h_{\ell}(t)\delta(\tau-\tau_{\ell})
\end{equation}
where $L$ is the total number of taps, $\{h_{\ell}(t)\}_{\ell=0,1,\cdots,L-1}$ the independent and Rayleigh distributed complex channel gains, $\{\tau_{\ell}\}_{\ell=0,1,\cdots,L-1}$ the timing delay of each path, and $\delta(\cdot)$ the delta function. Here, we assume that $\tau_{\ell}=\ell T_{s}$.

Up-converting $s(t)$ to a carrier frequency $f_{T}$, the post-channel equivalent signal takes the form
\begin{equation}
y(t)=\left[s(t)e^{j2\pi f_{T}t}\ast h(t,\tau)\right]+w(t)
\end{equation}    
where the notation $\ast$ stands for linear convolution, and $w(t)$ the complex, identically independently distributed (i.i.d.), additive white Gaussian noise (AWGN) with zero mean and variance $\sigma_{W}^{2}$; also, it is wide sense stationary (WSS), with independent real and imaginary part, and equal variance in both parts ($\sigma_{W}^{2}/2$). Now, assuming a CFO $\Delta f$ and a SFO $\eta$ given as
\begin{align}
\Delta f&\triangleq f_{T}-f_{R}\\
\eta&\triangleq (T_{s}'-T_{s})/T_{s}
\end{align}  
where $f_{R}$ is the deviated carrier frequency and $T_{s}'$ the deviated sampling interval at the receiver. The received $n$-th sample in the $l$-th OFDM block is 
\begin{equation}
r_{l,n}=y(t)e^{-j2\pi f_{R}t}\big|_{t=lN_{B}T_{s}'+NgT_{s}'+nT_{s}'},\ n=0,1,\cdots,N-1
\end{equation}    
After discarding the $N_{g}$ samples in the guard interval, the complex data for the $l$-th block and on the $k$-th subcarrier is \cite{Speth}
\begin{align}\label{equ:Rlk}
R_{l,k}&=X_{l,k}H_{l,k}\alpha(\Theta_{k})\left(e^{j\pi\Theta_{k}(N-1)/N}\cdot e^{j2\pi((lN_{B}+N_{g})/N)\Theta_{k}}\right)\nonumber\\
&+\mathrm{ICI}_{l,k}+W_{l,k}
\end{align} 
where $H_{l,k}$ is the channel transfer function (CTF) in frequency domain; $\Theta_{k}\approx \epsilon+\eta k$ and $\epsilon=\Delta f N_{B}T_{s}$ the normalized CFO to the subcarrier spacing; $\alpha(\Theta_{k})$ is the amplitude attenuation approaching unity and can be safely neglected; $\mathrm{ICI}_{l,k}$ the inter-carrier interference (ICI) due to distorted orthogonality of subcarriers; $W_{l,k}$ is the WSS i.i.d. Gaussian noise in frequency domain with independent real and imaginary parts. Without loss of generality, $\epsilon$ and $\eta$ can be regarded as the \emph{residual} part of CFO and SFO after coarse synchronization or imperfect channel estimation and equalization.  

\section{Proposed Technique}\label{sec:proposed}
\begin{figure}[t]
\centering
\includegraphics[width=20pc]{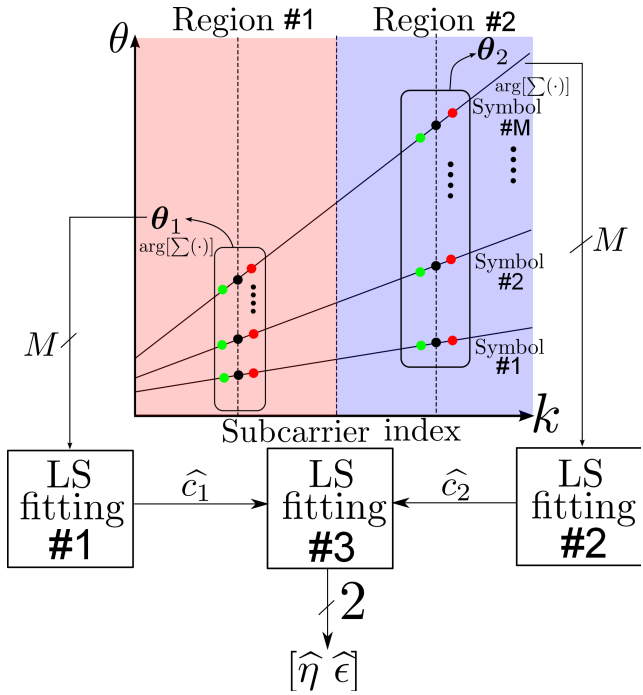}
\caption{The framework of the proposed technique. $\theta$ denotes the extra phase rotation caused by CFO and SFO. $Q=2$. The red dot and green dot form a pairwise correlation pair.}
\label{fig:structure}
\end{figure}

Define the full set of subcarrier index as $\mathbb{K}=\{k|0\leq k \leq N-1\}$, which can be further divided into $Q$ equally-spaced regions (assuming even $N$ and $Q$, and $N$ is divisible by $Q$), denoted as $\mathbb{K}=\mathbb{K}_{1}\cup\mathbb{K}_{2}\cup\cdots\cup\mathbb{K}_{q}\cup\cdots\cup\mathbb{K}_{Q}$ where
\begin{equation}
	\mathbb{K}_{q}=\left\{k\Bigg|\frac{(q-1)N}{Q}\leq k< \frac{qN}{Q}, k\in\mathbb{Z}\right\}
\end{equation}
Here, $\mathbb{Z}$ denotes integers. Ignoring disturbances of ICI, using equation \eqref{equ:Rlk}, for the $q$-th segment in the $l$-th OFDM block, the pairwise correlation is 
\begin{align}
&V_{l,k_{1},k_{2}}^{q}=R_{l,k_{1}}R_{l,k_{2}}\Big|_{k_{1}\in\mathbb{K}_{q},\ k_{2}\in\mathbb{K}_{q}}^{k_{1}+k_{2}=N_{q}}\nonumber\\
&=\big\{\lambda^{l}_{k_{1},k_{2}}e^{j\theta_{l,\epsilon,\eta}^{q}}\big\}\longmapsto \mathrm{Useful\ Part}\nonumber\\
&+\big\{X_{l,k_{1}}H_{l,k_{1}}W_{l,k_{2}}e^{j\theta_{l,\epsilon,\eta,k_{1}}}+X_{l,k_{2}}H_{l,k_{2}}W_{l,k_{1}}e^{j\theta_{l,\epsilon,\eta,k_{2}}}\nonumber\\
&+W_{l,k_{1}}W_{l,k_{2}}\big\}\longmapsto \mathrm{Cross\ terms}\label{equ:V}
\end{align}
where
\begin{align}
N_{q}&=\frac{N+2N(q-1)}{Q},\ q=1,2,\cdots,Q\\
\lambda^{l}_{k_{1},k_{2}}&=X_{l,k_{1}}X_{l,k_{2}}H_{l,k_{1}}H_{l,k_{2}}\\
\theta_{l,\epsilon,\eta}^{q}&=(2\epsilon +\eta N_{q})\left[2\pi l(1+g)+2\pi g+\pi \frac{N-1}{N}\right]\label{equ:extraphase}\\
\theta_{l,\epsilon,\eta,k_{1}}&=\pi\Theta_{k_{1}}\frac{N-1}{N}+2\pi(\frac{lN_{B}+N_{g}}{N})\Theta_{k_{1}}\\
\theta_{l,\epsilon,\eta,k_{2}}&=\pi\Theta_{k_{2}}\frac{N-1}{N}+2\pi(\frac{lN_{B}+N_{g}}{N})\Theta_{k_{2}}
\end{align}
and $g=N_{g}/N$. Clearly, the extra phase rotation of the useful part in \eqref{equ:extraphase} is irrelevant to subcarrier index $k_{1}$ and $k_{2}$; it is only pertinent to the OFDM block index $l$ and segment index $q$. The cross terms are the main disturbance in estimation. In practice, the contribution of signal and channel ($\lambda^{l}_{k_{1},k_{2}}$) should be replaced by  
\begin{equation}
\widehat{\lambda}^{l}_{k_{1},k_{2}}=\overline{X}_{l,k_{1}}\overline{X}_{l,k_{2}}\widehat{H}_{l,k_{1}}\widehat{H}_{l,k_{2}}
\end{equation}
where 
\begin{equation}\label{equ:X}
\overline{X}_{l,k}=\begin{cases}
X_{l,k},& k\in\mathbb{P}\\
\widehat{X}_{l,k},& k\in\mathbb{X}\\
0,& k\in\mathbb{U}
\end{cases} 
\end{equation}
The notation $\mathbb{P}$ is the full set of pilots and $\mathbb{U}$ the full set of null subcarriers. $\widehat{X}_{l,k}$ is the estimated $X_{l,k}$, obtained by the \emph{decision feedback device}. $\widehat{H}_{l,k}$ is the estimated CTF. $V_{l,k_{1},k_{2}}^{q}$ could be combined using a certain weight $\Gamma_{l,k_{1},k_{2}}$ given as
\begin{align}\label{equ:Gamma}
&\Gamma_{l,k_{1},k_{2}}=\nonumber\\
&\begin{cases}
\frac{1}{\sigma_{W}^{2}}\times\frac{1}{\left[|X_{l,k_{1}}|^{2}|H_{l,k_{1}}|^{2}+|X_{l,k_{2}}|^{2}|H_{l,k_{2}}|^{2}\right]+\sigma_{W}^{2}},&\mathrm{Weighted}\\
1,&\mathrm{Simplified}
\end{cases}
\end{align}
where $\sigma_{W}^{2}=\E\{|W_{l,k}|^{2}\}$. See Appendix \ref{sec:stat} for the details of such selection for the weighted $\Gamma_{l,k_{1},k_{2}}$. Computation of weighted $\Gamma_{l,k_{1},k_{2}}$ requires the second-order statistics of signal, channel and noise, avoided by the simplified scheme. For constant-modulus modulation, the weighted $\Gamma_{l,k_{1},k_{2}}$ reduces to
\begin{equation}
\Gamma_{l,k_{1},k_{2}}=\left(\left[|H_{l,k_{1}}|^{2}+|H_{l,k_{2}}|^{2}\right]\sigma_{S}^{2}\sigma_{W}^{2}+\sigma_{W}^{4}\right)^{-1}
\end{equation}
where $\sigma_{S}^{2}=\E\{|X_{l,k}|^{2}\}=\mathrm{Const.}$. To obtain estimation of $\theta_{l,\epsilon,\eta}^{q}$, we coherently stack $V_{l,k_{1},k_{2}}^{q}$ by 
\begin{equation}
Z_{l}^{q}=\sum_{(k_{1},k_{2})\in\mathcal{C}_{q}}V_{l,k_{1},k_{2}}^{q}\Gamma_{l,k_{1},k_{2}}[\widehat{\lambda}^{l}_{k_{1},k_{2}}]^{*},\ q=1,2,\cdots,Q
\end{equation}
where $(\cdot)^{*}$ denotes the conjugation of its argument and 
\begin{align}
\mathcal{C}_{q}&=\{(k_{1},k_{2})|k_{1}\in\mathbb{K}_{q,+},\ k_{2}\in\mathbb{K}_{q,-},\ k_{1}+k_{2}=N_{q}\}\nonumber\\
&\cap \{(k_{1},k_{2})|k_{1}\in \mathbb{U}^{c},k_{2}\in \mathbb{U}^{c}\}
\end{align}
$\mathbb{K}_{q,+}$ is the left half of $\mathbb{K}_{q}$ while $\mathbb{K}_{q,-}$ the right one; $\mathbb{U}^{c}$ is the \emph{absolute complement} of $\mathbb{U}$ given by $\mathbb{K}\backslash\mathbb{U}=\mathbb{P}\cup\mathbb{X}$; in general, the two-dimensional set $\mathcal{C}_{q}\neq\varnothing$. $\widehat{\theta}_{l,\epsilon,\eta}^{q}$ can be estimated by 
\begin{equation}
\widehat{\theta}_{l,\epsilon,\eta}^{q}=\arg\{Z_{l}^{q}\},\ q=1,2,\cdots,Q\label{equ:theta1l}
\end{equation}
The $M\times 1$ vector $\boldsymbol\theta_{q}=[\widehat{\theta}_{0,\epsilon,\eta}^{q}\ \widehat{\theta}_{1,\epsilon,\eta}^{q}\ \cdots\ \widehat{\theta}_{M-1,\epsilon,\eta}^{q}]^{T}$ can be linearized into
\begin{equation}
\boldsymbol\theta_{q}=\mathbf{A}\mathbf{b}_{q}+\boldsymbol\chi_{q}\label{equ:theta1}\\
\end{equation}
where $\mathbf{A}$ is the $M \times 2$ \emph{observation matrix} expressed by
\begin{align}
\mathbf{A}&=\begin{bmatrix}
D_{0} & D_{1} & \cdots & D_{l} & \cdots & D_{M-1}\\
V_{0} & V_{1} & \cdots & V_{l} & \cdots & V_{M-1}
\end{bmatrix}^{T}\\
D_{l}&=\pi(1+g)l\\
V_{l}&=2\pi(1+g)+\pi\left(\frac{N-1}{N}\right)
\end{align}
The $2\times 1$ vectors $\mathbf{b}_{q}$ take the form
\begin{equation}
\mathbf{b}_{q}=[c_{q}\ c_{q}]^{T}\label{equ:simpleb1}
\end{equation}
where \begin{math}c_{q}=2\epsilon+\eta N_{q}
\end{math};
$\boldsymbol\chi_{q}$ is the associated estimation error vector. By least square fitting, $\mathbf{b}_{q}$ is given by 
\begin{equation}
\widehat{\mathbf{b}_{q}}=\left(\mathbf{A}^{T}\mathbf{A}\right)^{-1}\mathbf{A}^{T}\boldsymbol\theta_{q}\label{equ:b1}
\end{equation}
Note that both $[\widehat{\mathbf{b}_{q}}]_{1,1}$ and $[\widehat{\mathbf{b}_{q}}]_{2,1}$ give estimation of $\widehat{c_{q}}$; $[\cdot]_{i,j}$ denotes the $(i,j)$-th entry of a vector/matrix. Here, we choose $[\widehat{\mathbf{b}_{q}}]_{1,1}$ and arrange all $\widehat{c_{q}}$ into the $Q\times 1$ vector $\mathbf{c}=[\widehat{c_{1}}\ \widehat{c_{2}}\ \cdots\ \widehat{c_{q}}\ \cdots \widehat{c_{Q}}]^{T}$ which leads to
\begin{equation}
	\mathbf{c}=\mathbf{B}\boldsymbol\mu+\boldsymbol\chi_{c}
\end{equation}
where $\boldsymbol\mu=[\eta\ \epsilon]^{T}$, $\boldsymbol\chi_{c}$ is the $Q\times 1$ error vector, and 
\begin{equation}
\mathbf{B}=\begin{bmatrix}
\frac{N}{Q} & \frac{3N}{Q} & \cdots & N_{q} & \cdots & \frac{N+2N(Q-1)}{Q}\\
2	    & 2	  	   & \cdots & 2     & \cdots & 2
\end{bmatrix}^{T}
\end{equation}
is the $Q\times 2$ observation matrix. Another least square fitting yields
\begin{equation}\label{equ:mu}
	\widehat{\boldsymbol\mu}=\left(\mathbf{B}^{T}\mathbf{B}\right)^{-1}\mathbf{B}^{T}\mathbf{c}
\end{equation}
The estimated $\widehat{\eta}$ and $\widehat{\epsilon}$ are $[\widehat{\boldsymbol\mu}]_{1,1}$ and $[\widehat{\boldsymbol\mu}]_{1,2}$ respectively. A simple sketch with $Q=2$ is drawn in Fig.~\ref{fig:structure}. 
Assuming correctness in tackling the phase ambiguity in the linearization process, derived in Appendix \ref{sec:stat}, estimation using either the weighted or simplified $\Gamma_{l,k_{1},k_{2}}$ is unbiased. On the other hand, the numerical variances of $\widehat{\eta}$ and $\widehat{\epsilon}$ are
\begin{align}
\Var\{\widehat{\eta}\}&=\frac{81\sum_{q=1}^{Q}U_{q}\left[\sum_{l=0}^{M-1}\frac{(2l-M+1)^{2}}{F_{l,q}}\right]}{32N^{4}(Q^{2}-1)^{2}\pi^{2}(1+g)^{2}M^{2}(M^{2}-1)^{2}}\label{equ:varf1}\\
\Var\{\widehat{\epsilon}\}&=\frac{81\sum_{q=1}^{Q}Y_{q}\left[\sum_{l=0}^{M-1}\frac{(2l-M+1)^{2}}{F_{l,q}}\right]}{32N^{4}(Q^{2}-1)^{2}\pi^{2}(1+g)^{2}M^{2}(M^{2}-1)^{2}}\label{equ:varf2}
\end{align}
where
\begin{align}
F_{l,q}=&\begin{cases}
\sum_{(k_{1},k_{2})\in\mathcal{C}_{q}}\frac{\phi_{l,k_{1},k_{2}}^{\times}}{\phi_{l,k_{1},k_{2}}^{+}+1},&\mathrm{Weighted}\\
\frac{(\sum_{(k_{1},k_{2})\in\mathcal{C}_{q}}\phi_{l,k_{1},k_{2}}^{\times})^{2}}{\sum_{(k_{1},k_{2})\in\mathcal{C}_{q}}\phi_{l,k_{1},k_{2}}^{\times}(\phi_{l,k_{1},k_{2}}^{+}+1)},&\mathrm{Simplified}
\end{cases}\label{equ:Fi}\\
\phi_{l,k_{1},k_{2}}^{\times}&=\frac{|X_{l,k_{1}}|^{2}|H_{l,k_{1}}|^{2}|X_{l,k_{2}}|^{2}|H_{l,k_{2}}|^{2}}{\sigma_{W}^{4}}\\
\phi_{l,k_{1},k_{2}}^{+}&=\frac{|X_{l,k_{1}}|^{2}|H_{l,k_{1}}|^{2}+|X_{l,k_{2}}|^{2}|H_{l,k_{2}}|^{2}}{\sigma_{W}^{2}}\\
U_{q}&=16N^{2}\left[2q-Q-1\right]^{2}\\
Y_{q}&=4N^{4}\left[2q-1-\frac{4Q^{2}-1}{3Q}\right]^{2}
\end{align}
Appendix \ref{sec:stat} validates $F_{l,q}|_{\mathrm{Weighted}}\geq F_{l,q}|_{\mathrm{Simp.}}$ and thus $\Var\{\widehat{\eta}\}|_{\mathrm{Weighted}}\leq \Var\{\widehat{\eta}\}|_{\mathrm{Simp.}}$, $\Var\{\widehat{\epsilon}\}|_{\mathrm{Weighted}}\leq \Var\{\widehat{\epsilon}\}|_{\mathrm{Simp.}}$. The equality establishes \emph{if and only if (iff)}
\begin{description}
\item[A1:] the channel experiences flat fading ($|H_{l,k}|^{2}\equiv 1/N$)
\item[A2:] constant modulus modulation ($|X_{l,k}|^{2}\equiv \sigma_{S}^{2}$)
\end{description}
Otherwise, weighted estimation always outperforms simplified estimation. To achieve the best performance, further assuming
\begin{description}
\item[A3:] equal and maximal \emph{cardinality} of each set $\mathcal{C}_{q}$, denoted by $N_{\mathcal{C}_{q}}=N/2Q$  
\end{description}
The variances under conditions A1$\sim$A3 are
\begin{align}
\Var\{\widehat{\eta}\}&=\frac{18Q^{2}}{\pi^{2}(1+g)^{2}M(M+1)(M-1)N^{3}(Q^{2}-1)\mathrm{SNR}}\label{equ:varetaa1a2}\\
\Var\{\widehat{\epsilon}\}&=\frac{6(4Q^{2}-1)}{4\pi^{2}(1+g)^{2}M(M+1)(M-1)N(Q^{2}-1)\mathrm{SNR}}\label{equ:varepsa1a2}
\end{align}
where $\mathrm{SNR}\triangleq \sigma_{S}^{2}\sigma_{H}^{2}/\sigma_{W}^{2}$; $\sigma_{H}^{2}=\E\{|H_{l,k}|^{2}\}$.\\
\textit{Remarks:}\\
\indent i) Apparently, an immediate way to enhance the performance is to raise $M$, which leads to asymptotically decreasing variances in cubic scale. Nevertheless, if the application is \textit{real-time oriented} rather than \textit{quality preferred}, where $\eta$ and $\epsilon$ should be tracked in the fastest manner, $M$ and $Q$ should be replaced by their minimums as $\min\{M\}=2, \min\{Q\}=2$.\\
\indent ii) If $[\widehat{\mathbf{b}_{q}}]_{2,1}$ is used for estimation, \eqref{equ:varetaa1a2} and \eqref{equ:varepsa1a2} are rewritten into 
\begin{align}
\Var\{\widehat{\eta}\}'&=\frac{(8M-4)(M-1)(1+g)^{2}}{(1+2g)^{2}}\Var\{\widehat{\eta}\}\label{equ:varetaa1a2alternative}\\
\Var\{\widehat{\epsilon}\}'&=\frac{(8M-4)(M-1)(1+g)^{2}}{(1+2g)^{2}}\Var\{\widehat{\epsilon}\}\label{equ:varepsa1a2alternative}
\end{align}
which are significantly higher than \eqref{equ:varetaa1a2} and \eqref{equ:varepsa1a2} increasing squarely with $M$. Therefore, it is reasonable to use $[\widehat{\mathbf{b}_{q}}]_{1,1}$.\\
\indent iii) According to \eqref{equ:V}, statistically, the proposed technique only relies on the independency and equal variance of the real and imaginary parts, and WSS assumptions of noise; it does not require the power spectrum density (PSD) of noise to be strictly flat (white), since the expectation of the cross terms is zero.         
 
\section{Simulation}\label{sec:sim}
In this section, we consider a wireless OFDM system with FFT size $N=512$. The guard interval is $N_{g}=64$. Thus, the length of an entire OFDM block is $N_{B}=576$. The total number of OFDM blocks is $M=10$, and the total number of segments is $Q=4$. The carrier frequency is set at $5$ GHz. The sampling period is $T_{s}=100ns$. For brevity and to exploit the best performance of the proposed estimators as well as other conventional pilot-assisted schemes, all subcarriers are regarded as pilots; otherwise, notations in \eqref{equ:X} must be used which varies with the accuracy of decision feedback device. The signal is modulated from 16-PSK constellation. The channel consists of $L=32$ Rayleigh taps, which are statistically independent distributed with a power delay profile decaying exponentially:
\begin{equation}
	\E\{|h_{\ell}|^{2}\}=\frac{\exp(-\ell/L)}{\sum_{\ell=0}^{L-1}|\exp(-\ell/L)|^{2}},\ \ell=0,1,2,\cdots,L-1
\end{equation}
For the proposed estimators and the scheme in \cite{JLS}, CTF is assumed to be known perfectly as well as $\sigma_{W}^{2}$\footnote{For OFDM systems containing null subcarriers, $\sigma_{W}^{2}$ could be estimated, which is omitted in this paper.}, unless otherwise mentioned. Mean squared error (MSE) results are used to benchmark the performance, defined as $\mathrm{MSE}\{\widehat{\eta}\}=\E\{|\widehat{\eta}-\eta|^{2}\}$ and $\mathrm{MSE}\{\widehat{\epsilon}\}=\E\{|\widehat{\epsilon}-\epsilon|^{2}\}$ where $\E\{\cdot\}$ denotes the expectation of its argument.
\begin{figure*}[tbp]
\begin{minipage}[t]{0.5\textwidth}
\centering
\includegraphics[width=21pc]{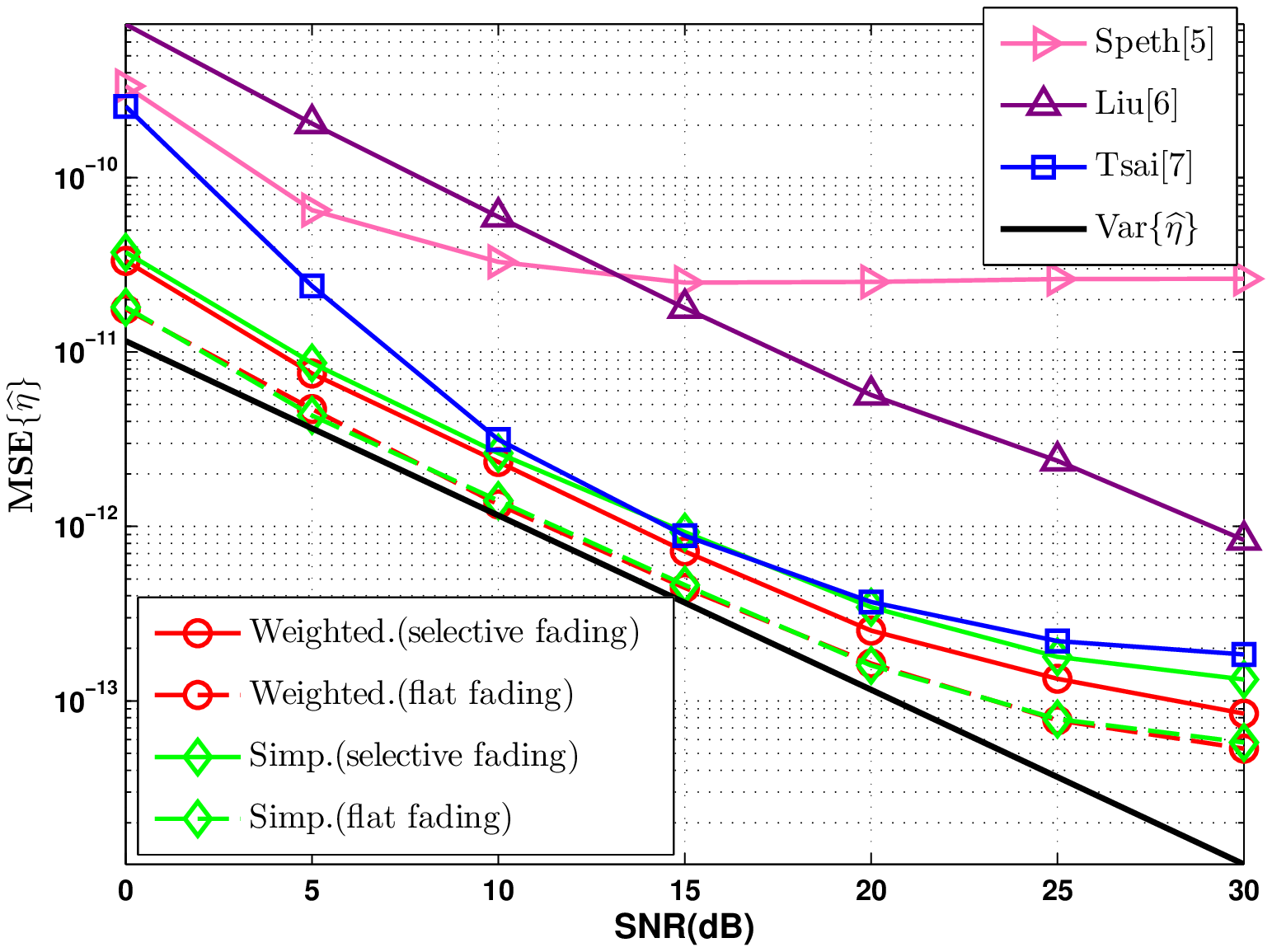}
\caption{$\mathrm{MSE}\{\widehat{\eta}\}$ comparison among different estimators; $M=10, Q=4, \eta=5\times 10^{-5}, \epsilon=0.02$, and $\mathrm{SNR}\in [0,30]$ dB.}
\label{fig:1}
\end{minipage}
\begin{minipage}[t]{0.5\textwidth}
\includegraphics[width=21pc]{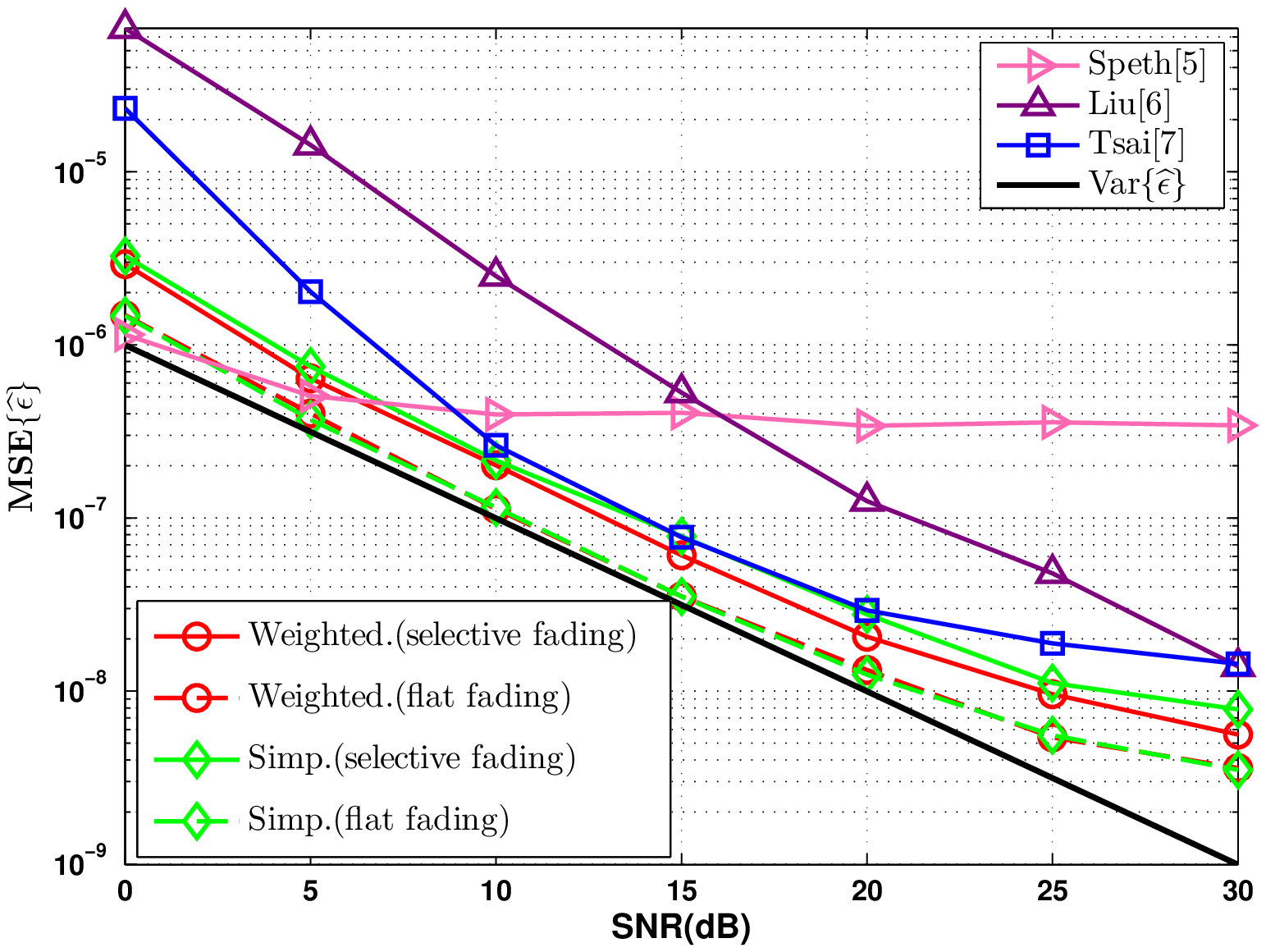}
\caption{$\mathrm{MSE}\{\widehat{\epsilon}\}$ comparison among different estimators; $M=10, Q=4, \eta=5\times 10^{-5}, \epsilon=0.02$, and $\mathrm{SNR}\in [0,30]$ dB.}
\label{fig:2}
\end{minipage}
\end{figure*}
\subsection{Comparison of $\mathrm{MSE}\{\widehat{\eta}\}$}
Fig.~\ref{fig:1} highlights the comparison of $\mathrm{MSE}\{\widehat{\eta}\}$ among the proposed estimators with other schemes in \cite{SpethII,ShouYinLiu,JLS}. Numerical result of $\Var\{\widehat{\eta}\}$ in \eqref{equ:varetaa1a2} is drawn by assuming A1$\sim$A3. In the multipath channel, both of the weighted and simplified estimator achieve the best performances. In the flat fading scenario, the weighted estimator reduces to the simplified one. $\Var\{\widehat{\eta}\}$ provides a tight bound in moderate $\mathrm{SNR}$.
\subsection{Comparison of $\mathrm{MSE}\{\widehat{\epsilon}\}$}
Fig.~\ref{fig:2} shows the performance comparison of $\mathrm{MSE}\{\widehat{\epsilon}\}$ in multipath channel. \cite{SpethII} could achieve the best performance when $\mathrm{SNR}\leq 7$ dB, which cannot be sustained into higher $\mathrm{SNR}$. Again, $\Var\{\widehat{\epsilon}\}$ provides a tight bound in moderate $\mathrm{SNR}$.  
\subsection{$\mathrm{MSE}\{\widehat{\eta}\}$ with a Varying $\eta$}
\begin{figure*}[tbp]
\begin{minipage}[t]{0.5\textwidth}
\centering
\includegraphics[width=21pc]{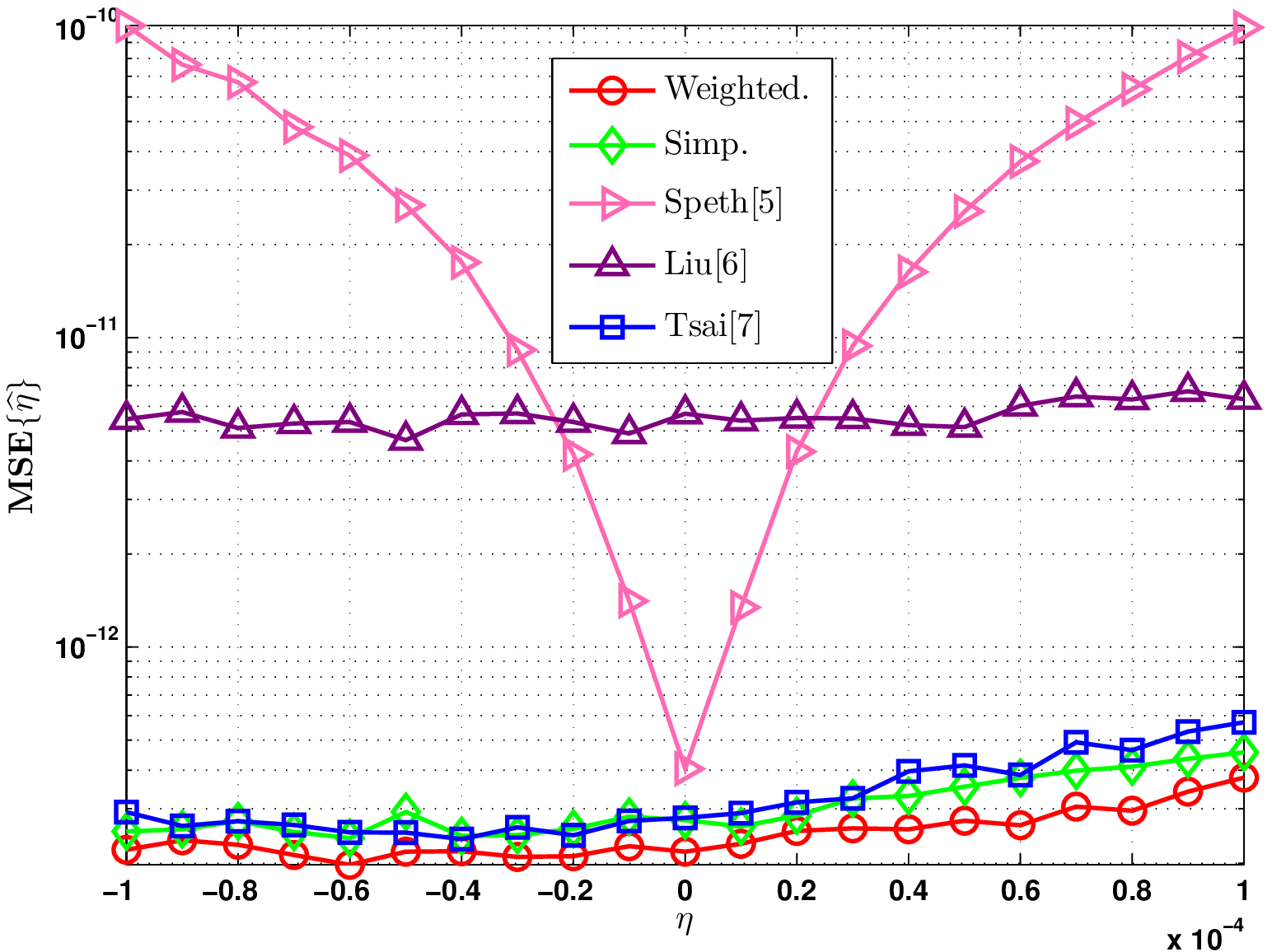}
\caption{$\mathrm{MSE}\{\widehat{\eta}\}$ under different $\eta$; $M=10, Q=4, \eta\in[-10:10]\times 10^{-5}, \epsilon=0.02$, and $\mathrm{SNR}=20$ dB.}
\label{fig:3}
\end{minipage}
\begin{minipage}[t]{0.5\textwidth}
\centering
\includegraphics[width=21pc]{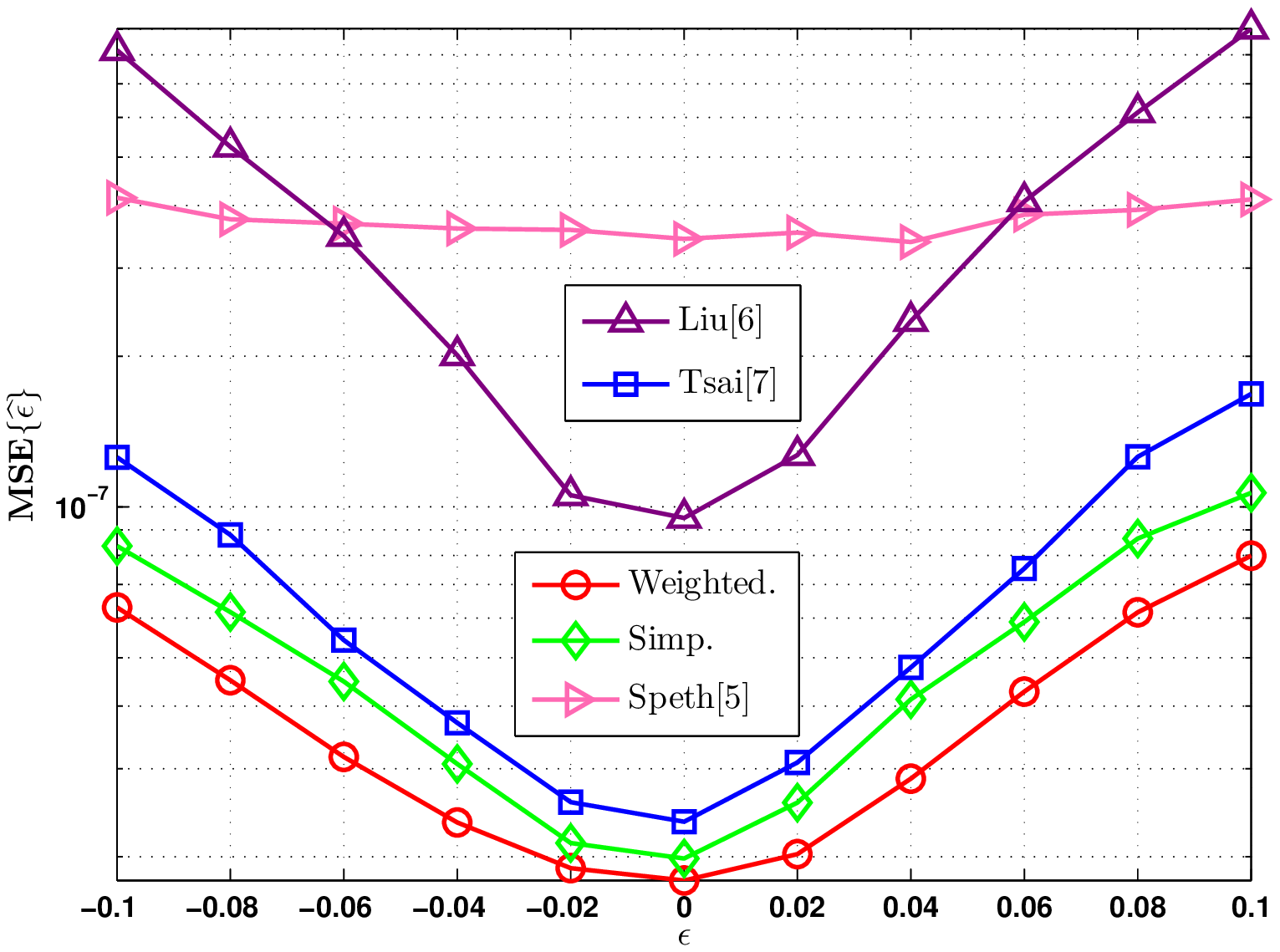}
\caption{$\mathrm{MSE}\{\widehat{\epsilon}\}$ under different $\epsilon$; $M=10, Q=4, \eta=5\times 10^{-5}, \epsilon\in[-0.1:0.1]$, and $\mathrm{SNR}=20$ dB.}
\label{fig:4}
\end{minipage}
\end{figure*}

Fig.~\ref{fig:3} shows the deviation of $\mathrm{MSE}\{\widehat{\eta}\}$ when $\eta$ changes under $\mathrm{SNR}=20$ dB in multipath channel. For the proposed estimators, $\mathrm{MSE}\{\widehat{\eta}\}$ is asymmetric for negative and positive $\eta$ due to the presence of a positive $\epsilon$; for $\eta>0$, the performance degrades gradually with a higher $\eta$ since the ICI is increasing simultaneously. For a major part, \cite{JLS} and the simplified estimator entangle with each other.
\subsection{$\mathrm{MSE}\{\widehat{\epsilon}\}$ with a Varying $\epsilon$}
Fig.~\ref{fig:4} displays the deviation of $\mathrm{MSE}\{\widehat{\epsilon}\}$ with a varying $\epsilon$ under $\mathrm{SNR}=20$ dB in multipath channel. Different from Fig.~\ref{fig:3}, shape of $\mathrm{MSE}\{\widehat{\epsilon}\}$ is akin to symmetric, since comparing with $\epsilon$, the contribution of $\eta$ on ICI is minor. In the full range, both of the proposed schemes outperform others.    
\begin{figure*}[tbp]
\begin{minipage}[t]{0.5\textwidth}
\centering
\includegraphics[width=21pc]{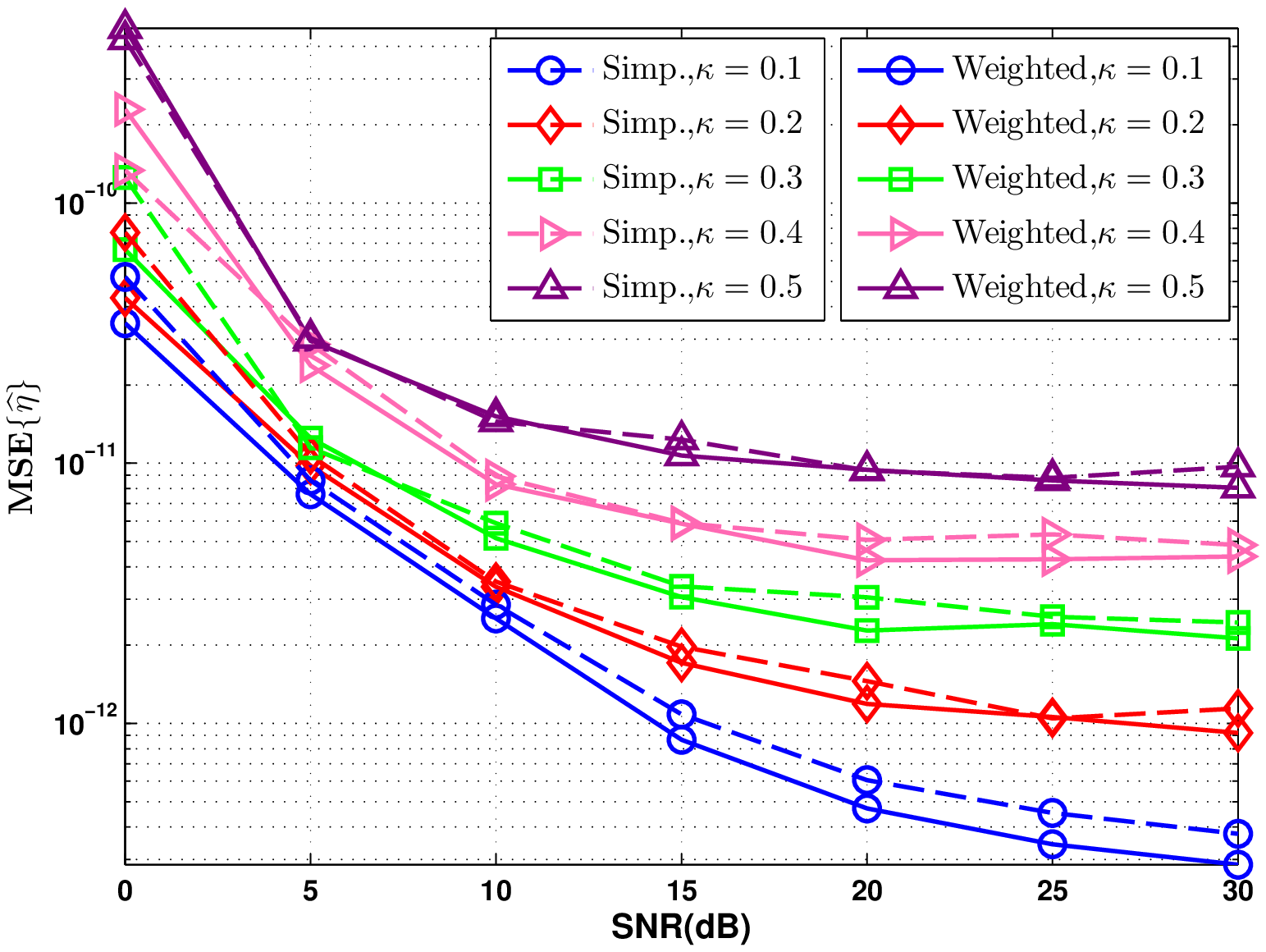}
\caption{$\mathrm{MSE}\{\widehat{\eta}\}$ under different accuracy of channel estimation; $\kappa\in[0.1:0.5], M=10, Q=4, \eta=5\times 10^{-5}, \epsilon=0.02$, and $\mathrm{SNR}\in [0,30]$ dB.}
\label{fig:5}
\end{minipage}
\begin{minipage}[t]{0.5\textwidth}
\centering
\includegraphics[width=21pc]{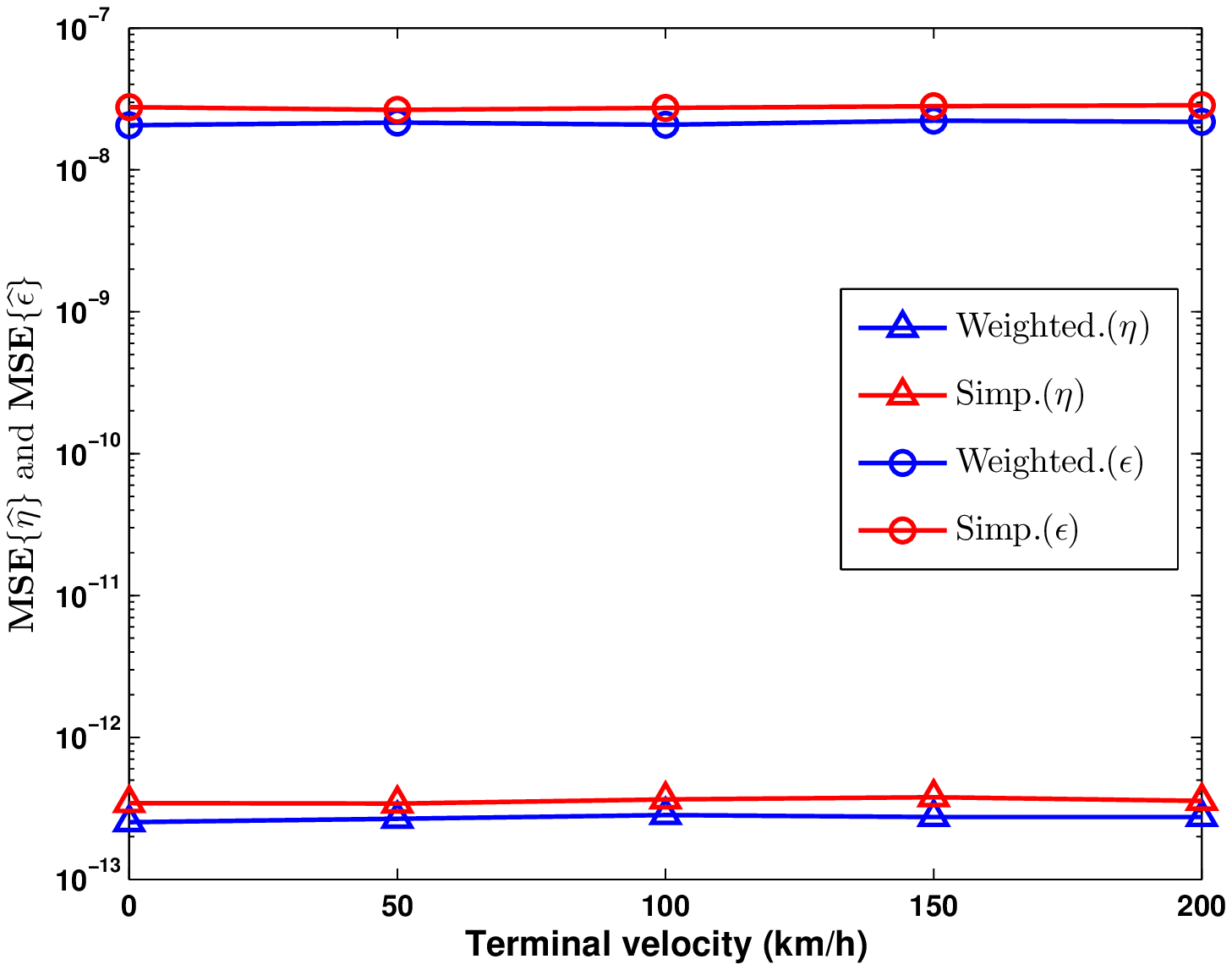}
\caption{$\mathrm{MSE}\{\widehat{\eta}\}$ and $\mathrm{MSE}\{\widehat{\epsilon}\}$ under different terminal velocity; $M=10, Q=4, \eta=5\times 10^{-5}, \epsilon=0.02$, and $\mathrm{SNR}=20$ dB.}
\label{fig:6}
\end{minipage}
\end{figure*}
\subsection{$\mathrm{MSE}\{\widehat{\eta}\}$ with Different Estimation Accuracy}
Fig.~\ref{fig:5} shows the $\mathrm{MSE}\{\widehat{\eta}\}$ performance with different channel estimation accuracy in multipath channel. Employing the same philosophy of \cite{Kyung}, the inaccurate channel estimation takes the form 
\begin{equation}
	\widetilde{H}_{l,k}=\sqrt{1-\kappa^{2}}H_{l,k}+\kappa J_{l,k}
\end{equation}
where $\kappa$ represents the estimation accuracy, and $J_{l,k}$ the additional complex noise with zero mean and variance $0.5$ in its real and imaginary part respectively, independent from $H_{l,k}$. The weighted estimator degrades significantly when severe inaccuracy occurs, which hinders the performance improvements especially in moderate to high $\mathrm{SNR}$ region. For moderate to high $\mathrm{SNR}$, the weighted estimator outperforms the simplified one under different $\kappa$. Similar conclusion can be drawn for $\mathrm{MSE}\{\widehat{\epsilon}\}$.   
\subsection{$\mathrm{MSE}\{\widehat{\eta}\}$ and $\mathrm{MSE}\{\widehat{\epsilon}\}$ under Mobility} 
Fig.~\ref{fig:6} exhibits the $\mathrm{MSE}\{\widehat{\eta}\}$ and $\mathrm{MSE}\{\widehat{\epsilon}\}$ in presence of terminal mobility with $\mathrm{SNR}=20$ dB under multipath channel. Merely the CSI pertinent to the first moment of the Rayleigh fading channel in simulation ($t$=0) is assumed to be known a priori; for the ensuing frames, the same CSI is used which entails a loss in channel estimation accuracy. The Doppler bandwidths with respect to terminal speed of $[50,100,150,200]km/h$ are $232,463,695,927$ Hz. In general, the performance deviation is insignificant if not imperceptible even the terminal velocity reaches $200km/h$, since the maximal value of the product between the Doppler bandwidth and the duration of an OFDM block is $5.3\times 10^{-2}$, a relatively small value. Thus, the CSI is sound enough to secure an excellent estimation.        

\section{Concluding Remarks}\label{sec:conclusion}
In this paper, we propose a joint estimation technique to deal with residual CFO and SFO estimation. By dividing the subcarrier index into a number of regions and exploiting the pairwise correlation, we estimate the phase increment between adjacent OFDM blocks, which yields accurate estimation after $Q+1$ times of least square fitting. Extensive simulations indicate better performance over several conventional pilot-assisted schemes. 

\renewcommand{\theequation}{\thesection.\arabic{equation}}
\appendices
\section{Bias and Variance of Estimation}\label{sec:stat}
First of all, consider the case of the simplified $\Gamma_{l,k_{1},k_{2}}$ in \eqref{equ:Gamma} and the statistical information of $\boldsymbol\theta_{q}$ in \eqref{equ:theta1l}. For the $l$-th element of $\boldsymbol\theta_{q}$ ($\widehat{\theta}_{l,\epsilon,\eta}^{q}$), subtracting the left hand side of \eqref{equ:theta1l} yields
\begin{equation}
\Delta \theta_{l,\epsilon,\eta}^{q}=\widehat{\theta}_{l,\epsilon,\eta}^{q}-\theta_{l,\epsilon,\eta}^{q}=\arg[Z_{l}^{q}/\exp(j\theta_{l,\epsilon,\eta}^{q})]
\end{equation}      
If $\widehat{\theta}_{l,\epsilon,\eta}^{q}$ is in vicinity of ${\theta}_{l,\epsilon,\eta}^{q}$, using the approximation $\tan(x)\approx x$ for $x$ small enough, we may write 
\begin{equation}\label{equ:vicinity}
\tan(\Delta \theta_{l,\epsilon,\eta}^{q})\approx \frac{\Im\{\Delta \theta_{l,\epsilon,\eta}^{q}\}}{\Re\{\Delta \theta_{l,\epsilon,\eta}^{q}\}} \approx \Delta \theta_{l,\epsilon,\eta}^{q}
\end{equation}
where $\Re\{\cdot\}$ and $\Im\{\cdot\}$ represent the real and imaginary part of the arguments. Expectation of $\Delta \theta_{l,\epsilon,\eta}^{q}$ in \eqref{equ:vicinity} is
\begin{equation}\label{equ:DeltaFrac}
\E\left\{\Delta \theta_{l,\epsilon,\eta}^{q}\right\} \approx \E \left\{\frac{\Im\{\Delta \theta_{l,\epsilon,\eta}^{q}\}}{\Re\{\Delta \theta_{l,\epsilon,\eta}^{q}\}}\right\} \approx \frac{\E\{\Im\{\Delta \theta_{l,\epsilon,\eta}^{q}\}\}}{\E\{\Re\{\Delta \theta_{l,\epsilon,\eta}^{q}\}\}}
\end{equation}
which holds if \begin{math}
\E\{\Re[\Delta \theta_{l,\epsilon,\eta}^{q}]\} \gg [\Var\{\Re[\Delta \theta_{l,\epsilon,\eta}^{q}]\}]^{1/2}
\end{math}. In fact, each component in $\Im\{\Delta \theta_{l,\epsilon,\eta}^{q}\}$ contains either $W_{l,k_{1}}$, $W_{l,k_{2}}$, or $W_{l,k_{1}}W_{l,k_{2}}$ (see \eqref{equ:V}) and therefore, $\E\{\Im\{\Delta \theta_{l,\epsilon,\eta}^{q}\}\}=0$, which finally leads to the unbiasedness of $\widehat{\boldsymbol\mu}$ in \eqref{equ:mu} since only linear intermediate operations are involved. 

For the numerical variance of $\Delta \theta_{l,\epsilon,\eta}^{q}$, we could use
\begin{equation}\label{equ:Varequ}
\Var[\Delta \theta_{l,\epsilon,\eta}^{q}]\approx\frac{\Var[\Im\{\Delta\theta_{l,\epsilon,\eta}^{q}\}]}{(\E[\Re\{\Delta \theta_{l,\epsilon,\eta}^{q}\}])^{2}} 
\end{equation}
if \begin{math}
\E\{\Re[\Delta \theta_{l,\epsilon,\eta}^{q}]\} \gg [\Var\{\Re[\Delta \theta_{l,\epsilon,\eta}^{q}]\}]^{1/2}
\end{math}. Standard calculations yield

\begin{align}
&\Var[\Im\{\Delta\theta_{l,\epsilon,\eta}^{q}\}]=\sum_{k_{1},k_{2}}\sigma_{W}^{2}|H_{l,k_{1}}|^{2}|H_{l,k_{2}}|^{2}\times\nonumber\\
&\left[|X_{l,k_{1}}|^{2}|H_{l,k_{1}}|^{2}+|X_{l,k_{2}}|^{2}|H_{l,k_{2}}|^{2}+\sigma_{W}^{2}\right]\\
&(\E[\Re\{\Delta \theta_{l,\epsilon,\eta}^{q}\}])^{2}=\bigg\{\sum_{k_{1},k_{2}}|X_{l,k_{1}}|^{2}|X_{l,k_{2}}|^{2}|H_{l,k_{1}}|^{2}|H_{l,k_{2}}|^{2}\bigg\}^{2}
\end{align}
Note that, for visual clearance, we abbreviate the notation $(k_{1},k_{2})\in \mathcal{C}_{q}$ with $k_{1},k_{2}$. Thus,
\begin{equation}
\Var[\Delta \theta_{l,\epsilon,\eta}^{q}]\approx F_{l,q}|_{\mathrm{Simp.}}
\end{equation}
where $F_{l,q}$ is defined in \eqref{equ:Fi}. Using the linear intermediate manipulations, we derive \eqref{equ:varf1} and \eqref{equ:varf2}. Substituting $F_{l,q}|_{\mathrm{Simp.}}$ with $F_{l,q}|_{\mathrm{Weighted.}}$ backward produces the weighted version of $\Gamma_{l,k_{1},k_{2}}$ in \eqref{equ:Gamma} after lengthy calculations. 

To prove $F_{l,q}|_{\mathrm{Weighted}}\geq F_{l,q}|_{\mathrm{Simp.}}$, we invoke the Cauchy-Schwarz-Inequality \cite{CauchySchwarz}. The essential steps are listed below.
\begin{proof}\label{proof:CSI}
To prove
\begin{equation}
\sum_{k_{1},k_{2}}\frac{\phi_{l,k_{1},k_{2}}^{\times}}{\phi_{l,k_{1},k_{2}}^{+}+1}\geq \frac{(\sum_{k_{1},k_{2}}\phi_{l,k_{1},k_{2}}^{\times})^{2}}{\sum_{k_{1},k_{2}}\phi_{l,k_{1},k_{2}}^{\times}(\phi_{l,k_{1},k_{2}}^{+}+1)}
\end{equation}
is equivalent to prove
\begin{equation}\label{equ:prove1}
\sum_{k_{1},k_{2}}\frac{\phi_{l,k_{1},k_{2}}^{\times}}{\phi_{l,k_{1},k_{2}}^{+}+1}\sum_{k_{1},k_{2}}\phi_{l,k_{1},k_{2}}^{\times}(\phi_{l,k_{1},k_{2}}^{+}+1)\geq (\sum_{k_{1},k_{2}}\phi_{l,k_{1},k_{2}}^{\times})^{2}
\end{equation}
Now, using
\begin{math}
A(k_{1},k_{2})=\frac{\phi_{l,k_{1},k_{2}}^{\times}}{\phi_{l,k_{1},k_{2}}^{+}+1}
\end{math}
and 
\begin{math}
B(k_{1},k_{2})=\sum_{k_{1},k_{2}}\phi_{l,k_{1},k_{2}}^{\times}(\phi_{l,k_{1},k_{2}}^{+}+1)
\end{math}, 
the Cauchy-Schwarz-Inequality gives
\begin{equation}\label{equ:prove2}
	\sum_{k_{1},k_{2}}A(k_{1},k_{2})\sum_{k_{1},k_{2}}B(k_{1},k_{2})\geq\left\{\sum_{k_{1},k_{2}}\sqrt{A(k_{1},k_{2})B(k_{1},k_{2})}\right\}^{2}
\end{equation}
where the right hand side of \eqref{equ:prove2} is \emph{exactly} the right hand side of \eqref{equ:prove1}. The '$=$' holds iff    
\begin{equation}\label{equ:suffcond}
\phi_{l,k_{1},k_{2}}^{+}=\mathrm{Const.},\ \forall (k_{1},k_{2})\in \mathcal{C}_{q}
\end{equation}
and we have the conditions A1, A2 \footnote{Actually, it is merely a sufficient condition of \eqref{equ:suffcond}. Specifically, consider that there is a \emph{sage} at the transmitter side who could render $|X_{l,k}|^{2}=C|H_{l,k}|^{-2},\ \forall k$ where C is a constant, then \eqref{equ:suffcond} also establishes. However, it is not insightful to be pursued.}. Therefore, we verify that $F_{l,q}|_{\mathrm{Weighted}}\geq F_{l,q}|_{\mathrm{Simp.}}$.
\end{proof}

\bibliography{Residual} 
\end{document}